%% file: sn-article.tex
\documentclass[sn-mathphys-num]{sn-jnl}

\usepackage{graphicx}%
\usepackage{multirow}%
\usepackage{amsmath,amssymb,amsfonts}%
\usepackage{amsthm}%
\usepackage{mathrsfs}%
\usepackage[title]{appendix}%
\usepackage{xcolor}%
\usepackage{textcomp}%
\usepackage{manyfoot}%
\usepackage{booktabs}%
\usepackage{algorithm}%
\usepackage{algorithmicx}%
\usepackage{algpseudocode}%
\usepackage{listings}%
\usepackage{url}
\usepackage{soul}

\begin{document}

\title[Article Title]{Swift: High-Performance Sparse Tensor Contraction for Scientific Applications}

\author{\fnm{Andrew} \sur{Ensinger}}\email{ensingea@oregonstate.edu}

\author{\fnm{Gabriel} \sur{Kulp}}\email{kulpga@oregonstate.edu}

\author{\fnm{Victor} \sur{Agostinelli}}\email{agostinv@oregonstate.edu}

\author{\fnm{Dennis} \sur{Lyakhov}}\email{lyakhovd@oregonstate.edu}

\author{
    Lizhong Chen\footnote{Corresponding author. Email: chenliz@oregonstate.edu, Telephone: (541) 737-3317, Fax: (541) 737-1300}
}

\affil{\orgdiv{Electrical Engineering and Computer Science}, \orgname{Oregon State University}, \orgaddress{\street{1500 SW Jefferson Way}, \city{Corvallis}, \postcode{97331}, \state{Oregon}, \country{United States}}}

\newcommand\this[0]{Swift}
\abstract{
In scientific fields such as quantum computing, physics, chemistry, and machine learning, high dimensional data are typically represented using sparse tensors. Tensor contraction is a popular operation on tensors to exploit meaning or alter the input tensors. Tensor contraction is, however, computationally expensive and grows quadratically with the number of elements. For this reason, specialized algorithms have been created to only operate on the nonzero elements. Current sparse tensor contraction algorithms utilize sub-optimal data structures that perform unnecessary computations which increase execution time and the overall time complexity. We propose \textit{\this{}}, a novel algorithm for sparse tensor contraction that replaces the costly sorting with more efficient grouping, utilizes better data structures to represent tensors, and employs more memory-friendly hash table implementation.
\this{} is evaluated against the state-of-the-art sparse tensor contraction algorithm, demonstrating up to 20$\times$ speedup in various test cases and being able to handle imbalanced input tensors significantly better.
}




\maketitle

\section{Introduction}
Tensors are multidimensional arrays which are used to represent data in the fields of quantum chemistry, quantum physics, machine learning, and more\cite{nwchem,Fishman,Koppl,Lingjie,Pavosevic,Chase}. In many scenarios it is common for the tensors used in these fields to have a small percentage of nonzero elements, in which case they are referred to as \textit{sparse tensors} \cite{Choy_2020}. Tensor contraction (matrix multiplication in higher dimensions) is an operation that can be used to exploit meaning or alter the input tensors to better understand the data they represent. However, this operation is computationally expensive \cite{kisil2021} and execution time increases quadratically with respect to the number of elements contained in the input tensors. This can slow down the progress of research which is why computer scientists have created specialized software algorithms and hardware modules specifically to perform the operation of sparse tensor contraction. Sparse tensor contraction is the contraction of two tensors where one or both of them are sparse, which can be performed orders of magnitude faster than traditional tensor contraction algorithms on sparse input tensors because the speed of the operation scales as a function of the nonzero elements rather than the shape.

Conventional tensor contraction algorithms do not take advantage of sparsity which results in redundant calculations being performed with zero-valued entries. Specialized algorithms \cite{sparta,Xiao2024,Xiao2022,athena} attempt to overcome this by ignoring entries that have no value by doing element-wise operations on the remaining important elements. The main challenges facing sparse tensor contraction algorithms are: irregular memory access patterns, unknown output tensor size, and high amounts of intermediate data. Irregular memory access patterns arise from the fact that in sparse tensor contraction, elements are typically fetched from memory based on their coordinates in the tensor. The coordinate of a tensor element can be broken up into the contracting modes (i.e., dimensions) which are used to find what elements it should contract with, and free modes which are used to determine where the contracted element is located in the output tensor. Unknown output tensor size arises from the idea that even if it is known the density of the input tensors, it is impossible to know the density of the output tensor. This is because tensor contraction is dependent on \textit{where} those elements are located in the input tensors which would be unknown without processing all the elements. Finally, high amounts of intermediate data arise from the fact that after the output entries are calculated there will be a large portion of them which share the same coordinates. These entries must be found and added together to have one output element representing that coordinate value. 

Sparse tensor contraction hardware accelerators, including Tensaurus\cite{tensaurus}, Extensor\cite{extensor}, and Outerspace\cite{outerspace}, are time consuming to physically design and are much more expensive compared to a software implementation. Consequently, high performance software contraction algorithms are naturally attractive, as they can be employed for noteworthy speedups over conventional software libraries but without requiring additional hardware.

Consequently, several popular machine learning software libraries, such as PyTorch \cite{pytorch2019} and TensorFlow \cite{tensorflow2016}, offer support for sparse tensor contraction specifically. However, due to the main challenges of sparse tensor contraction, there are limited high performance software algorithms. The most prominent is called the Sparta algorithm \cite{sparta} which can be summarized into 4 main stages. The first stage involves sorting the input tensor $\mathcal{X}$ by free mode(s) and hashing the input tensor $\mathcal{Y}$ by contracting mode(s). This will make the input tensors easier to process. The second step is the contraction stage which involves fetching elements from the two input tensors that share the same contracting modes, and saving an output entry which has its coordinates located at the concatenation of the free modes of both entries and its value is the product of the two input entries. The third step is to find entries which share the same coordinates and add their values together so only one entry remains to represent the value at that coordinate. Finally, the fourth step is to write all of the output elements to a data structure which represents the output tensor. Sparta also offers an optional fifth step to sort the output entries but it is unnecessary and won't be focused on in this work. 

There are three areas of possible improvement of this algorithm which this work aims to address.
First, input tensor $\mathcal{X}$ needs to have input entries that share the same free mode next to each other in memory. The use of sorting to achieve this goal is a very computationally expensive choice which increases the execution time and time complexity. Second, the method of using a hash table to efficiently fetch elements from an input tensor based on contracting mode is generally cache-unfriendly\cite{hashtable2022} and cause unnecessary searching through other entries which were hashed to the same slot. Third, Sparta's use of chained hash tables leads to irregular memory access patterns, unnecessarily bloating memory access time. 

We propose \textit{\this{}}, an improved algorithm for sparse tensor contraction that replaces the costly sorting with more efficient grouping, utilizes better data structures to represent tensors, and employs more memory-friendly hash table implementation. The choice of data structures is guided on what would experience the most cache-efficiency and how well it integrates into the following stages in the algorithm. 
For the input processing stage, our algorithm incorporates grouping, which ensures input tensor elements with the same free modes are located next to each other in memory, to both improve the execution time and time complexity compared to sorting the entries. This technique makes the time complexity of the input processing stage $\mathcal{O}({n})$ with respect to the number of entries in both input tensors, which is an improvement upon the $\mathcal{O}({n}log(n))$ time complexity of the state-of-the-art input processing stage. In our scheme, the second input tensor is represented as an array rather than a hash table which improves memory access time as elements are located next to each other in memory to utilize cache-locality, rather than using pointers to locate elements. Finally, for the accumulation stage a probing hash table is used to locate elements with the same contracting modes compared to a chaining hash table which leads to pointer chasing (following chains of pointers to distant locations in memory). The performance benefits of these changes are evaluated using the contraction time and peak memory consumption relative to the state-of-the-art sparse tensor contraction algorithm. A  variety of different test cases were used that vary in entry count, order, and evenness of entries in both tensors. 

The main contributions of this work are: 
\begin{itemize}
  \item Analyzed state-of-the-art sparse tensor contraction algorithm and inefficiencies. 
  \item Proposed an improved algorithm with explanation of key aspects to achieve better performance while improving time complexity. 
  \item Evaluated the proposed algorithm versus state-of-the-art sparse tensor contraction program for various input tests cases, demonstrating 2-20$\times$ speedup in execution time.
\end{itemize}

\section{Background \& Related Work}

\subsection{Sparse Tensor Contraction}
\subsubsection{Tensors} 
Tensors are multidimensional arrays which have found applications in numerous scientific fields that work with higher dimensional data \cite{cichocki2014}.  A \textit{mode} is another title for dimension, so an $N$-dimension tensor is alternatively known as an $N$-mode tensor. The number of modes a tensor has is also referred to as the \textit{order} of the tensor. For example, the temperature at any given point in a room could be represented as 3-mode tensor, which has order 3 because each point in the room could be addressed by 3D coordinates, and the temperature would be the value of the tensor at these coordinates.

It is standard notation to symbolize a tensor as $\mathcal{X}^{I_1\times I_2 \times ...\times I_n}$, and individual tensor entry values as $x_{i_1i_2...i_n}$. $I_k$ denotes the length of the $k$th mode of the tensor and $i_k$ denotes an index along this mode where $0 \leq i_k < I_k$. For example, tensor $\mathcal{X}^{6\times 3 \times 4}$ would have a maximum index of 5 in the first dimension, a maximum index of 2 in the second dimension, and a maximum index of 3 in the third dimension. $x_{4,2,3}$ would denote the value of the tensor entry located at $(4,2,3)$. 

\subsubsection{Sparse Tensors} 
\begin{figure}[hbt!]
	\centering
	\includegraphics[width=0.85\textwidth]{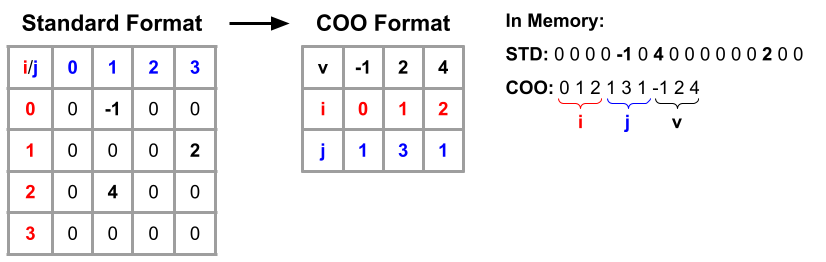}
	\caption{Example of converting standard tensor format to COO format in memory. i and j represent the rows and columns of the value denoted by v. }
	\label{fig:representation}
\end{figure}

Sparse tensors have a majority of entries with the value of 0, where the percentage of nonzero elements is referred to as the \textit{density}. Because of this, it would be inefficient to represent a sparse tensor with the standard method of listing the elements contiguously in memory, as the majority of the entries will have the value 0. Instead, the elements of a sparse tensor would be better represented using COOrdinate List format (COO) \cite{tensorflow2016}.  This methodology lists the elements as coordinate value pairs, ignoring pairs that have a value equal to zero. This ensures that sparse tensors are represented in such a way that only meaningful elements are accounted for. An example of this can be seen in Figure \ref{fig:representation}. In this example the density is $\frac{3}{16} = 18.75\%$, because there are 3 nonzero elements ($NNZ$) and 16 total elements. In practical applications the density of sparse tensors can range from $0.01\%$ to $10\%$ \cite{feasta2024,smith2015,jiajia2016,jiajia2018,jiajia2019,jiajia2020} which leads to much greater memory savings. 

 \subsection{Tensor contraction} 
 Tensor contraction can be expressed using the form shown in Equation \ref{eq:general}, which can be expanded to Equation \ref{eq:general_expanded}. 

  \begin{equation} \label{eq:general} 
 	\mathcal{Z} = \mathcal{X} \otimes_{\{C^Y\}}^{\{C^X\}} \mathcal{Y}  
\end{equation} 

\begin{equation} \label{eq:general_expanded}
	z_{\{f^X\}\{f^Y\}} = \sum_{\{c\}}^{} x_{\{f^X\}\{c\}} \cdot y_{\{c\}\{f^Y\}}
\end{equation}

The tensor contraction operation takes input tensor $\mathcal{X}$, input tensor $\mathcal{Y}$, the list of contracting $\mathcal{X}$ modes (top), and the list of contracting $\mathcal{Y}$ modes (bottom), as exemplified in Equations \ref{eq:example} \& \ref{eq:example_expanded}. If a certain mode of tensor $\mathcal{X}$ is contracted with another mode of tensor $\mathcal{Y}$, both of these modes must have the same length, similar to how for matrix multiplication the number of columns in the first matrix must match the number of rows in the second matrix. The tensor contraction operation focuses on finding entries in $\mathcal{X}$ and $\mathcal{Y}$ that share contracting indices, multiplying the values together, and adding that result to the output tensor $\mathcal{Z}$. This can be seen in Equations \ref{eq:general_expanded} and \ref{eq:example_expanded} where a matching contracting index is denoted as \textit{c}. All instances of matching contracting indices are summed with output coordinates which share the same free modes of $\mathcal{X}$ and $\mathcal{Y}$, hence the summation sign.

\begin{equation} \label{eq:example} 
 	\mathcal{Z} = \mathcal{X} \otimes_{\{1,3\}}^{\{1,2\}} \mathcal{Y}  
\end{equation} 

\begin{equation} \label{eq:example_expanded}
	z_{\{x_3,x_4,y_2,y_4\}} = \sum_{\{c\}}^{} x_{\{x_3,x_4\}\{c\}} \cdot y_{\{c\}\{y_2,y_4\}}
\end{equation}

\begin{figure}[hbt!]
	\centering
	\includegraphics[width=0.9\textwidth]{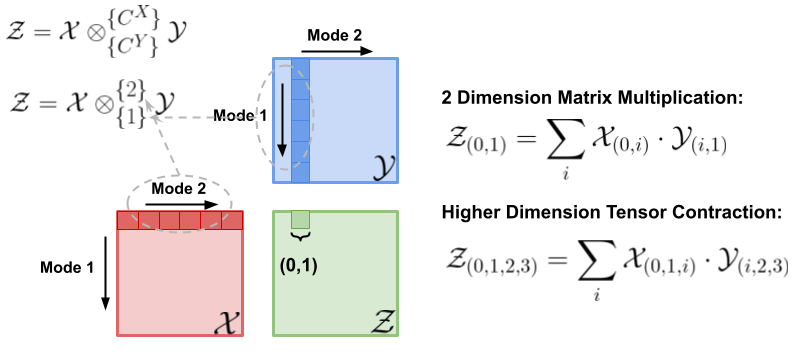}
	\caption{Example of tensor contraction with 2-dimensional input tensors. In this case the operation is a matrix multiplication where the second mode of $\mathcal{X}$ is contracted with the first mode of $\mathcal{Y}$. As shown the dimensionality of the problem can be increased by adding more indices to match the dimension of the application. }
	\label{fig:contraction_example}
\end{figure}

Tensor contraction is a generalization of matrix multiplication as shown in Figure \ref{fig:contraction_example}. Note that the element at index $(i,j)$ in the output tensor $\mathcal{Z}$ is the dot product of the $i$th row of tensor $\mathcal{X}$ with the $j$th column of tensor $\mathcal{Y}$. The resulting element in tensor $\mathcal{Z}$ is the concatenation of the free modes (first mode from $\mathcal{X}$ and second mode from $\mathcal{Y}$, which correspond to indices 0 and 1, respectively), which mirrors Equation \ref{eq:general_expanded}. The example shown in this figure can be applied to tensors with higher dimensionality which leads to the general concept of tensor contraction. Sparse tensor contraction occurs when one or both of the input tensors of a contraction operation are sparse \cite{Jiajia_2016}. 

\subsection{Hash Tables}
Hash tables are used repeatedly in both related work and our proposed algorithm. After the element-wise contraction stage, which will be explained in section 3, each of the output entries must be combined with entries that share the same coordinate. Hash tables are a common solution for this problem as they operate on key-value pairs. In this instance the key would be coordinate of the entry and the value would be the value of the entry. The key is mapped to a table slot using a hash function, and the key-value pair is placed at that slot. The hash table achieves $O(1)$ search time derived from the fact that this hashing process described is theoretically invariant of how many elements are currently in the table \cite{hashtable2022}. A major design decision with this data structure is how to handle the scenario of two different keys being hashed to the same slot. The two most prominent solutions are probing and chaining. 

\subsubsection{Linear Probing Hash Tables} Linear probing hash tables handle collision by placing the key-value pair at the next available slot in the table by iterating down the table until it finds an empty location. If the value needs to be fetched, the key will be hashed to the original slot and iterated down until the key is found and the value is returned. One obvious downside of this approach is that as the table begins to fill and probing will take longer. This is why it is a common practice to make probing hash tables $\times 1.3$ \cite{Cafiero_2023} the expected maximum number of elements that need to be stored. 

\subsubsection{Chaining Hash Tables} Chaining hash tables handle collision by treating each slot as if it were a linked list (i.e., the entire hash table is an array of linked lists) and adding the new entry by hashing to a slot and adding to the head of the linked list. If the value needs to be fetched, the key will hash to the correct slot and the linked list will be traversed until the desired key-value pair is found and the value will be returned. A disadvantage of this approach is that it leads to irregular memory accesses due to the pointer chasing involved in traversing a linked list which often consumes much more time than a probing hash table. A major advantage of a chaining hash table is that it can store an unknown number of elements because linked lists can grow indefinitely whereas the probing hash table is fixed in size. 

In practice, neither of these methods will achieve constant hash time because the more entries in the same slot the longer it takes to fetch or add a new element. However, near constant hashing time will be achieved, so long as the number of elements is comparable to the number of slots.

\subsection{Related Work}
\begin{figure}[hbt!]
	\centering
	\includegraphics[width=1\textwidth]{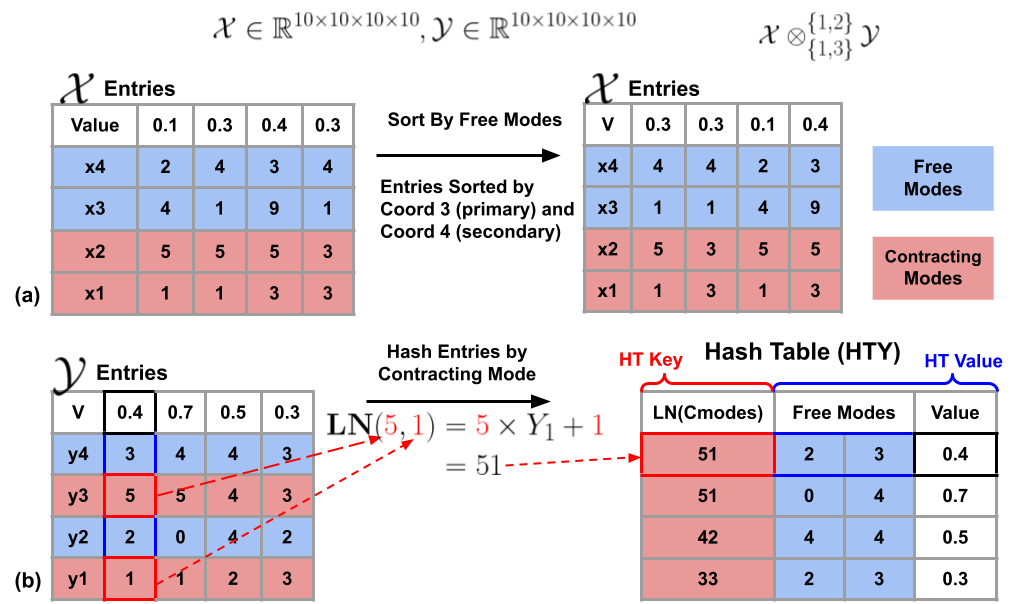}
	\caption{Visualization of Input Processing for Sparta Algorithm. (a) The X entries are sorted by the free modes and (b) the Y entries are hashed into a table where the key is the contracting mode and the value is the free mode with the corresponding entry value. Here, lower case letters (y1-y4) indicate indices of coordinates, and upper case letters ($Y_1$) indicates the length of the mode, which is 10 in this example. }
	\label{fig:sparta_preprocessing}
\end{figure}

\begin{figure}[hbt!]
	\centering
	\includegraphics[width=1\textwidth]{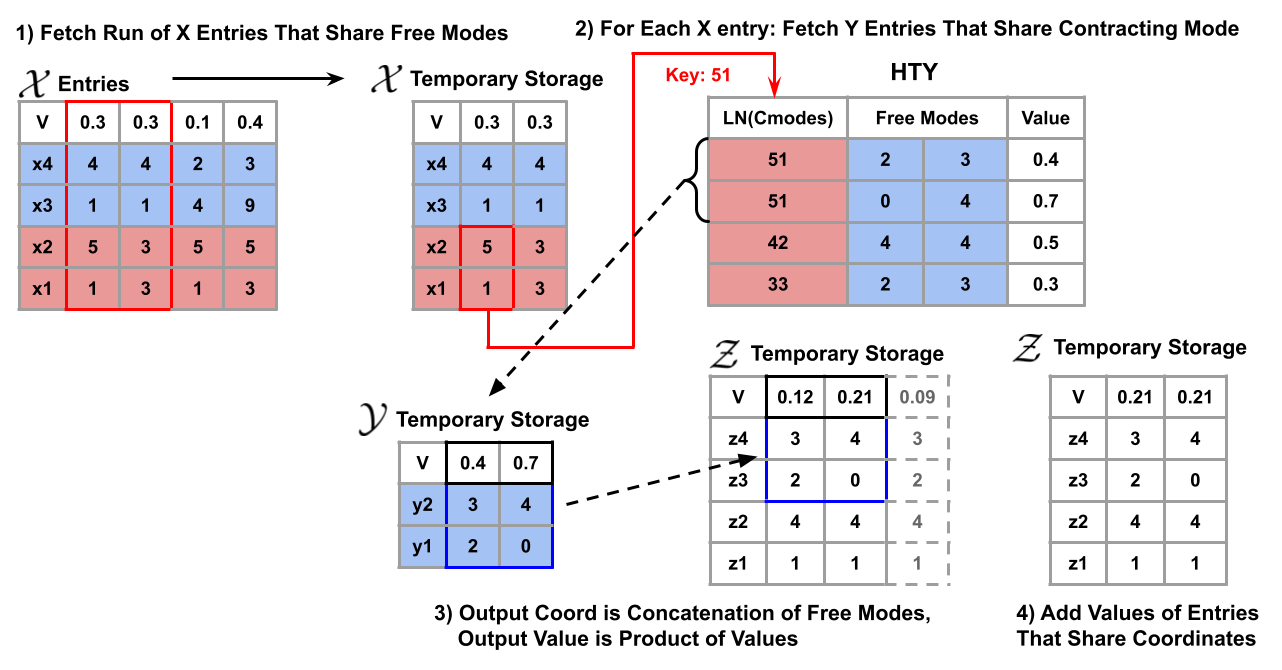}
	\caption{Visualization of Contraction and Accumulation process for Sparta Algorithm. If the contracting modes of an X entry match with the contracting modes of a Y entry, the output entry coordinate is the concatenation of free modes and the value is the multiplication of the individual values. }
	\label{fig:sparta_contraction}
\end{figure}

 Acceleration of sparse tensor contraction has been approached using hardware accelerators and high performance software algorithms. Hardware accelerators \cite{tensaurus,extensor,outerspace, flaash2024, matraptor2020} almost always outperform software algorithms, but there is extreme commitment in the creation of hardware.
 In the realm of high performance software, Sparta is a state-of-the-art sparse tensor contraction algorithm which operates on COO formatted tensors. This algorithm addresses the main concerns with sparse contraction which are: irregular memory access patterns, unknown output tensor size, and high amounts of intermediate data. Machine learning libraries such as PyTorch and TensorFlow offer support for sparse tensor contraction but they do not have comparable contraction time to Sparta. The Sparta algorithm can be broken down into the Input Processing stage (Figure \ref{fig:sparta_preprocessing}), Contraction \& Accumulation (Figure \ref{fig:sparta_contraction}), and Writeback, which are explained below.

\subsubsection{Input Processing} Input processing comprises of manipulating the input tensors $\mathcal{X}$ and $\mathcal{Y}$ such that they can be more easily processed by the following stages. An example of this is shown in Figure \ref{fig:sparta_preprocessing}. The objective is to have the entries of tensor $\mathcal{X}$ with the same free modes to be located next to each other for easier data fetching and contradiction (more in the next subsection). Thus the tensor entries will be sorted by free mode. In this examples, coordinates 3 and 4 make up the free modes for this tensor so the entries will be sorted by coordinate 3 primarily and coordinate 4 secondarily. 

For the contraction operation, the $\mathcal{Y}$ entries must be easily fetched from memory based on its contracting mode. To achieve this, Sparta converts tensor $\mathcal{Y}$ from a list of entries into a hash table representation which has O(1) access time given a search query. This is shown in Figure \ref{fig:sparta_preprocessing}(b) where the key is the large number representation (LN; see the figure for calculation illustration) of the contraction modes, and the values of the hash table is the free modes and the value of the tensor at that coordinate. As you can see in this example, there are 2 coordinates which share the same key (51) in the hash table. If this key were to be requested, both entries would be fetched from the table. For the sake of clarity we show these two entries in the table as separate but in implementation they would both be contained in the same slot. 

\subsubsection{Contraction \& Accumulation} 
Due to input sorting, a subset of $\mathcal{X}$ entries that share the same free mode can be fetched from memory together. Each $\mathcal{X}$ entry in this list will be iterated over during the contraction phase, and the process for each entry is as follows. First the contracting modes of the $\mathcal{X}$ will be noted. Next, every entry from tensor $\mathcal{Y}$ that has the same contracting mode will be fetched from the hash table. The coordinate of the output tensor is the concatenation of the free modes of tensor $\mathcal{X}$  with the free modes of tensor $\mathcal{Y}$. The value of the output entry is simply the product of the $\mathcal{X}$ value with the $\mathcal{Y}$ value. The output entry is then placed into an accumulation hash table. 

During accumulation, there will be a large number of intermediate output entries, many of which have duplicate coordinates. Entries with matching coordinates must be located efficiently so their values can be added together. Because output entry coordinates are a concatenation of the tensor $\mathcal{X}$ free mode and the tensor $\mathcal{Y}$ free mode, two output entries will certainly not have matching coordinates if the coordinate contribution from entry $\mathcal{X}$ is not the same. This is why it would be advantageous to sort the entries of tensor $\mathcal{X}$ in the input processing stage so that all of the intermediate output entries with matching $\mathcal{X}$ free modes are located next to each other. Because the entries with matching $\mathcal{X}$ coordinates are already next to each other in temporary storage, all that remains to be done is finding matching tensor $\mathcal{Y}$ coordinates to see if the overall output coordinates match. An example of this can be seen in Figure \ref{fig:sparta_contraction}. In this case the free mode being contracted is $(1,4)$ so all $\mathcal{X}$ entries that have this free mode pair are fetched from memory and placed into temporary storage. For each of the $\mathcal{X}$ entries in temporary storage, the hash table HtY is used to fetch all $\mathcal{Y}$ that share the same contracting mode. In this example, the first $\mathcal{X}$ entry is being contracted which has a contracting mode set of $(1,5)$ so all $\mathcal{Y}$ entries that share this contracting mode set are placed into temporary storage. Note that only the free modes must be saved for each $\mathcal{Y}$ entry fetched because the contracting modes of these entries are not present in the output tensor $\mathcal{Z}$entries. As previously stated, the output entries are a concatenation of the free modes of $\mathcal{X}$ and the free modes of $\mathcal{Y}$ so these $\mathcal{Y}$ free modes are concatenated with the free mode of $\mathcal{X}$ and added to $\mathcal{Z}$ temporary storage. All output entries that share the same coordinate are then mapped to the same slot and their values are added together. In this case, two output entries share the same coordinate $(1,4,2,3)$ so their values $0.12$ and $0.09$ are added to get the final value for this coordinate which is $0.21$. This Figure shows the accumulation happening after contraction for ease of understanding, but in the implementation entries are filtered as they enter the accumulation hash table for temporary storage of $\mathcal{Z}$ entries. 

\subsection{Writeback} After Contraction \& Accumulation, because the algorithm was implemented using parallelism, each thread has a sub-collection of entries which belong to the output tensor. During Writeback, the thread-local collections of entries are iterated over and saved to a global data structure which represents the output tensor. 

\section{Motivation}
\label{others}

\subsection{Time Complexity} 
Time complexity of Sparta can be calculated by adding the time complexities of each individual stage in order to give an overview of how well the algorithm scales. Equation \ref{eq:tc_sparta} shows the time complexity of the Sparta Algorithm where $nnz_X$ denotes the number of nonzero entries in tensor $\mathcal{X}$, and $cmode_{Yavg}$ denotes the average number of entries associated with each contracting mode of $\mathcal{Y}$. 

The first term expresses the time complexity of the input processing stage. The input processing stage has time complexity $\mathcal{O}(nlog(n))$ with respect to, $nnz_X$, the number of nonzero entries in tensor $\mathcal{X}$ because its entries were sorted using quicksort. In contrast, the input processing stage has linear time complexity with respect to, $nnz_Y$, the number of nonzero entries in tensor $\mathcal{Y}$ because each was hashed into a table. The second term is the time complexity of the contraction and accumulation phases which must iterate over each nonzero $\mathcal{X}$ entry, and for each $\mathcal{X}$ entry the average mode length of $\mathcal{Y}$ must be iterated over to get the resulting output entry which is added to a hash table. Finally, the  third term expresses the time complexity of the Writeback stage which is equal to the number of output entries because each entry must be saved to the output tensor. 

\begin{align}
	T_{Sparta} &=  T_{Processing} + T_{Contraction} + T_{Writeback} \\
    \text{where:}&\nonumber\\  
    \nonumber
    T_{Processing} &= \mathcal{O}({nnz}_Xlog(nnz_X) + nnz_Y) \\ \nonumber
    T_{Contraction} &= \mathcal{O}(nnz_X \times cmode_{Yavg}) \\ \nonumber
    T_{Writeback} &= \mathcal{O}(nnz_Z) 
\end{align} \label{eq:tc_sparta}

The time complexity of the Contraction \& Accumulation stages cannot be improved upon as it is necessary to iterate over each $\mathcal{X}$ entry and the corresponding $\mathcal{Y}$ entries to generate the correct output entries. Similarly, the time complexity of the Writeback stage cannot be improved upon as each entry must be exported. There is, however, potential to improve the time complexity of the input processing stage. It is also, of course, possible to improve the overall performance of each stage, even if the time complexity remains unchanged. Accessing the same number of elements sequentially in memory generally leads to performance improvement compared to random access, and two linear algorithms can have vastly different execution times based on the number of steps per element even if that number is constant. 

\subsection{Hash Table Leads to Increased Memory Access Time}
Sparta represents the tensor $\mathcal{Y}$ as a hash table (HtY) which can be used to fetch a run of entries that share the same contracting mode ($C^Y$). Although the creation of the hash table can be done in $O(n)$ time and is generally fast, there is room for improvement when it comes to the time it takes to fetch elements during the contraction stage of the algorithm. 

\begin{figure}[hbt!]
	\centering
	\includegraphics[width=1\textwidth]{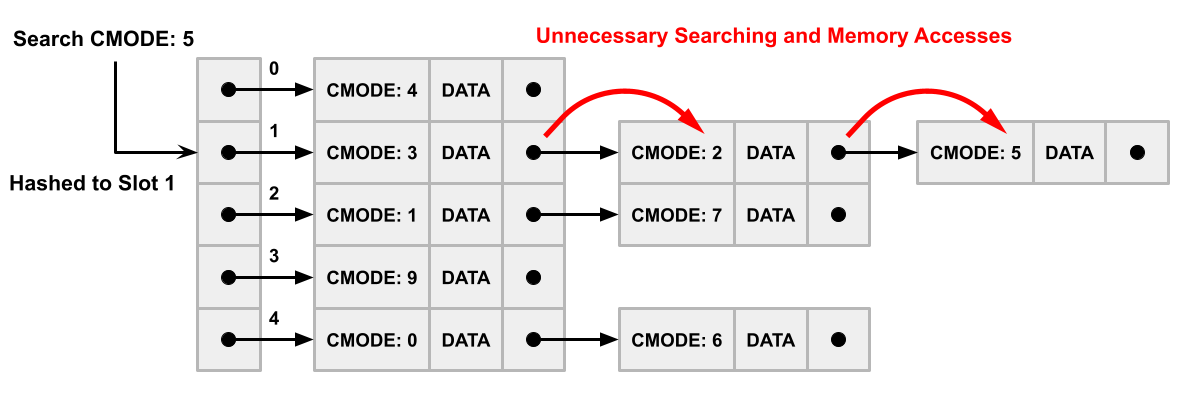}
	\caption{Example of hash table search requiring unnecessary pointer chasing. If the requested contracting mode was 5, the search would  cross contracting modes 3 and 2 which are a part of the same slot.}
	\label{fig:ht_problem}
\end{figure}

One aspect that makes this approach in Sparta slow compared to an array is the fact that chaining hash tables (the variant used in Sparta) leads to pointer chasing when traversing a linked list. This causes irregular memory accesses which harms cache performance. The second aspect that makes this approach slower is that during the hashing process the hash function commonly maps different keys to the same slot. This means if a contracting mode must be searched in the table, there will be undesired elements to filter through to access the desired entries. An example of this can be seen in Figure \ref{fig:ht_problem}. In this case the requested contracting mode is 5, shown in top left, but two other nodes must be traversed to access that element because 3 entries were mapped to that slot of the table. 

Combining the above two aspects, in a chaining implementation of a hash table, an additional pointer to the next entry must be fetched for each undesired contracting mode. This consumes a lot of time as there is a high probability a new line of memory must be fetched and placed into cache numerous times during this process.

Similarly, chaining hash tables are also used for the storage of output entries in the accumulation stage where an alternative data structure would have much better memory access time and cache-efficiency. 

\subsection{Input Sorting is Unnecessary for Input Processing}
As previously stated, the input processing stage is composed mainly of the sorting of the $\mathcal{X}$ entries and the hashing of the $\mathcal{Y}$ entries. Our experiments show that the sorting of $\mathcal{X}$ vastly overshadows the hashing of $\mathcal{Y}$ which is reflected in the time complexity equation of Sparta. During the input processing stage of Sparta, tensor $\mathcal{X}$ is sorted by the free modes of its coordinates in order to get matching free modes next to each other. Sorting also has the additional property of putting these entries in ascending order based on the free modes. If there is a way to remove the reliance on the ascending order while still maintaining the correctness of contraction, much time can be saved from not needing to sort $\mathcal{X}$. This is achieved by using our proposed methods described in the following section. 

\section{Approach} 

In this section, we present our proposed \textit{\this{}}, a novel and much improved algorithm for sparse tensor contraction. Our approach improves upon Sparta by replacing sorting with grouping, utilizes better data structures to represent tensors, and employs memory-friendly hash table implementation. In the following subsections, we describe the proposed improvement to the input processing, contraction, and accumulation stages in detail.

\subsection{Perform Grouping at Input Processing Stage}

If entries of tensor $\mathcal{X}$ are fetched in chunks where free modes are shared amongst entries, fetching is rendered more efficient due to locality. Therefore, the input processing stage should arrange the entries of tensor $\mathcal{X}$ such that entries which share free modes are placed contiguously in an array. This allows for a block of entries that share the same free mode to be fetched in constant time utilizing pointers to the start of runs. During the input processing stage, \this{} utilizes grouping by free mode instead of sorting by free mode to manipulate the order of the input entries. Grouping by free mode simply means placing entries that share free mode next to each other without ensuring that entry free modes are in any particular order. While this might still take $\mathcal{O}({n}log(n))$ in general cases, a linear time version is possible here by assigning a small metric (numerical proxy) to represent each unique free mode and applying a less stringent sorting algorithm to this metric. In particular, we employ a relaxed counting sort in this work, which is an algorithm that can achieve linear time complexity when the sorting metric does not exceed the number of entries being sorted so it is ideal for this scenario. 

\subsubsection{Counting Sort for Small Value Ranges} 
\begin{figure}[hbt!]
	\centering
	\includegraphics[width=1\textwidth]{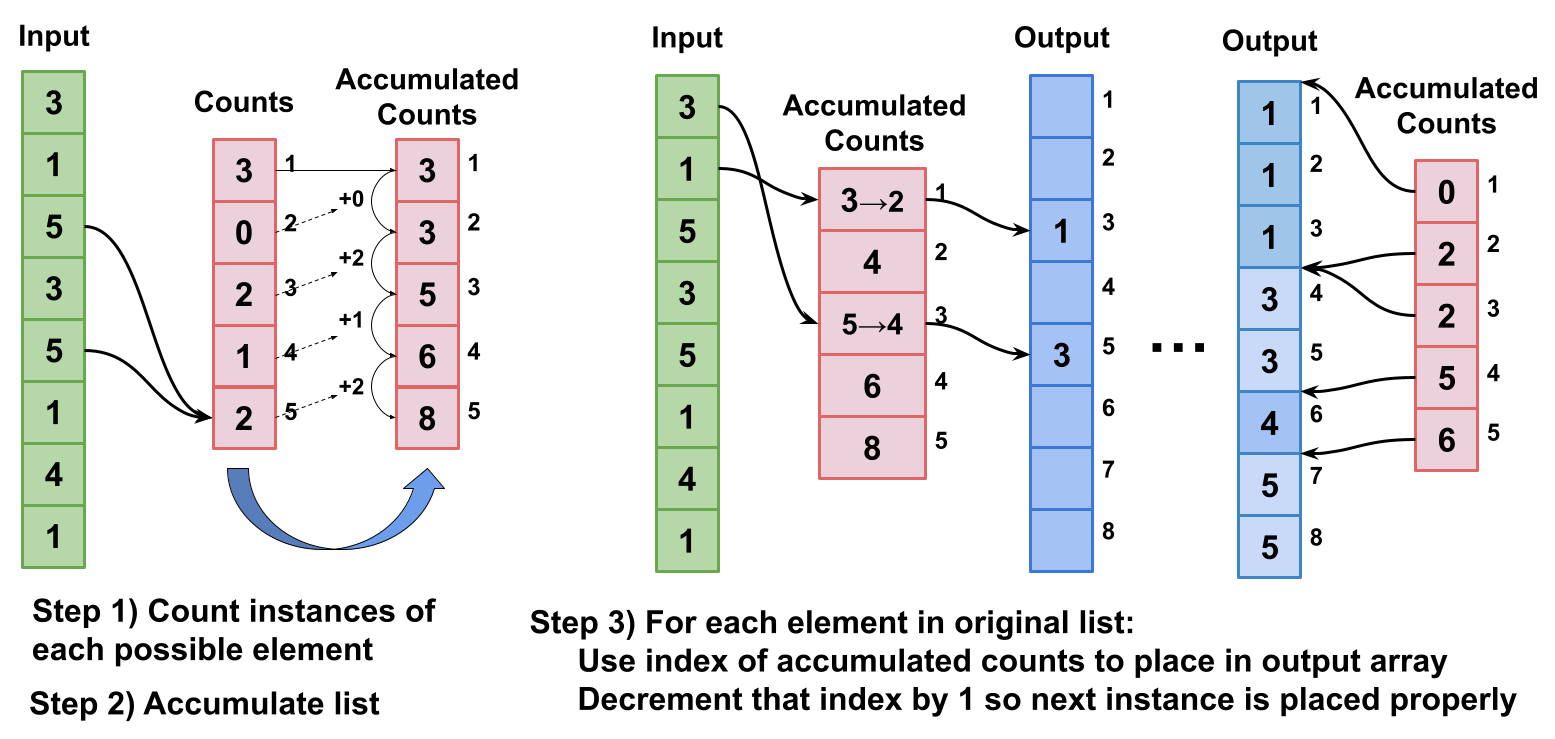}
	\caption{Example of counting sort applied to a list of integers with fixed range of elements from 1 to 5. Note that after this process is completed there is an array that conveniently points to divisions in elements which could be used for $O(1)$ search time. }
	\label{fig:counting_sort}
\end{figure}

Counting sort is a sorting algorithm that achieves a time complexity of $O(n + k)$ where $n$ is the number of elements and $k$ is the number of different values that elements can be assigned. An example of this algorithm can be seen in Figure \ref{fig:counting_sort}. In the first step, the array of elements is iterated over and the total number of instances of each possible value is found and placed into a count array (e.g., element 1 appears 3 times). The algorithm then takes the cumulative sum of the Count Array, yielding dividers to where elements of each value should be placed. Finally, the input array is iterated over and placed at the corresponding pointer (e.g., element 3 is placed at index 5 in the output), and the accumulated count is then decremented. The final output is a sorted array. This algorithm is much faster than quicksort when $k$ is small as it only requires 2 iterations over the input array and 1 iteration over the Count Array. If $k$ can be guaranteed to be less than or equal to $n$, this algorithm achieves $O(n)$ sorting.

\subsubsection{Small Numbers for Free Modes}
\begin{figure}[hbt!]
	\centering
	\includegraphics[width=1\textwidth]{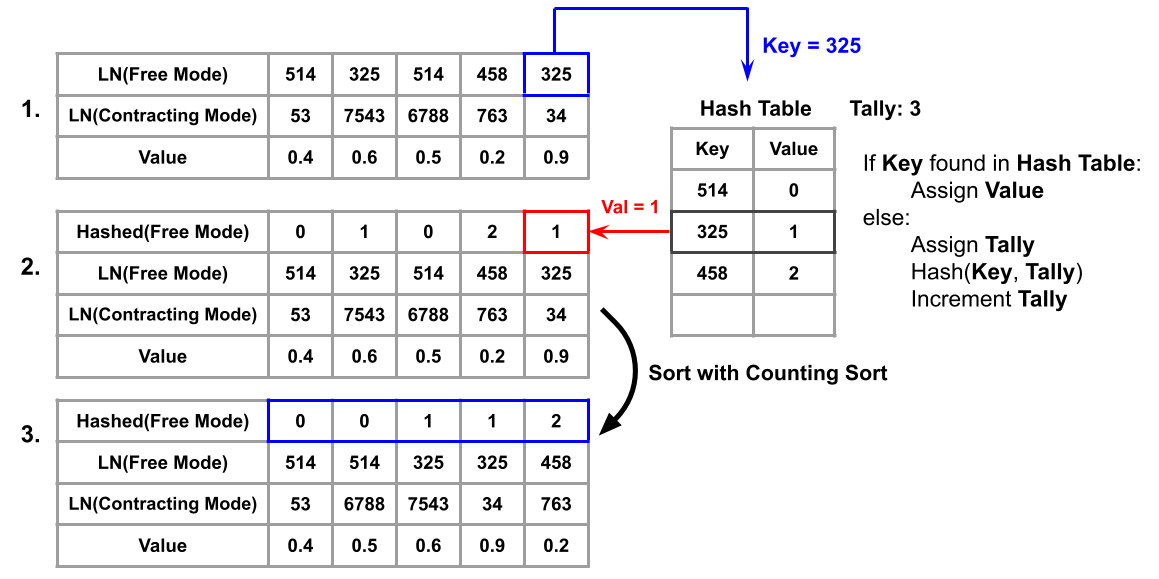}
	\caption{Grouping Process. First use hash table to replace all of the large number representations of contraction modes with a smaller mapping. Second, sort the values by contracting mode with counting sort. }
	\label{fig:grouping}
\end{figure}

In order to render counting sort more efficient than alternatives, the value of $k$ must be reduced, ideally to be less than $n$ for linear time complexity. Assignment of a metric with the smallest range of values can be done by hashing the free modes of each entry which is the process described in Figure \ref{fig:grouping}. 
The first step is to go through each entry and map the free modes representation to a metric which has a minimum range of possible values. This can be done by using a hash table to see if the free modes have already been hashed. If so, the entry is assigned the smaller metric. If not, the entry is assigned the metric equal to the number of elements currently in the table. Using the example shown in Figure \ref{fig:grouping}, the key $325$ was found in the hash table with the value $1$, thus that entry was assigned the metric $1$ in the second array. If the key was $400$, it would not be found in the hash table, and the entry would have been assigned metric $3$ because there are currently $3$ key value pairs in the hash Table. The reason for the hashing to smaller numbers is because Counting Sort has a time complexity dependent on the range of possible elements.

\subsubsection{Fast Grouping}\label{sec:fastgrouping}
The mapping of free modes to a smaller metric described in the previous subsection has guaranteed the maximum sorting metric is no greater than the number of elements in the array. Thus Counting Sort can be applied to this sorting metric which will have a time complexity of $\mathcal{O}(n + k)$, where $k \leq n$, which means grouping the elements has a time complexity of $\mathcal{O}(n)$. Note that after the sorting process the free modes are not in any particular order, i.e. ascending or descending, but that is of no importance for the contraction algorithm.  

\subsection{Array Representation of Tensor Y}
A similar process to the one described in Section \ref{sec:fastgrouping} above will be performed on the entries of tensor $\mathcal{Y}$ but with respect to the contracting modes instead of the free modes. The access time for a sequence of $\mathcal{Y}$ entries that share a desired contracting mode can be found in O(1) time. This can be done using the table of pointers to divisions in contracting modes which is given by Counting Sort results in an array of sorted elements with a corresponding vector of pointers to divisions of contracting modes, the access time for a sequence of $\mathcal{Y}$ entries that share a desired contracting mode can be found in O(1) time using the table of pointers to divisions in contracting modes. 

When it comes to implementation, this data structure is significantly faster than the hash tables used in Sparta for three reasons. First, with a hash table it is very likely that two blocks of contracting modes will be hashed to the same slot. As a consequence, in order to get the desired run of contracting indices, the program must unnecessarily filter through the extraneous contracting modes that were hashed to the same slot. Second, the entries are stored contiguously in memory which greatly improves cache efficiency. Third, the performance of a hash table can vary based on its implementation. In the case of a chaining hash table, excessive time may be spent following pointers, while a probing hash table aiming for peak execution time is expected to consume around 1.3 times more memory compared to an array. For HtY, Sparta implements a chaining hash table whereas we use the faster probing hash table at the cost of slightly increased memory.

\subsection{Probing Hash Table for Intermediate Entry Storage} 
Chaining tables are required for Sparta's accumulation stage because the $NNZ$ of the output tensor is unknown. 
However, the accumulation stage could be improved by using a probing hash table instead of a chaining hash table to store output entries from the contraction stage. Due to cache locality, it is much faster to iterate through a table in array format than to perform pointer chasing as would be done with a chaining table. \ul{The challenge lies in the fact that a probing hash table requires a predefined size, which can be difficult to estimate before the contraction process begins.} 

We address this challenge with the following clever method. During the grouping stage of tensor $\mathcal{Y}$, a vector is available that records the number of entries corresponding to each contracting index. One could go through every $\mathcal{X}$ entry that shares the same free mode and accumulate the number of $\mathcal{Y}$ entries associated with each of these operations. This can be used to assign an upper bound for the number of slots in the probing hash table such that dynamically allocated memory is not required. Using Figure \ref{fig:counting_sort} as an example, the contents of the array could represent the large number representation of the contracting modes of tensor $\mathcal{Y}$, thus the \textit{counts} array will store the count of each contracting mode set. 

As an example, given a set of 4 $\mathcal{X}$ entries that share the same free modes, the associated contracting modes could be $[1,2,3,3]$. Thus, the total number of $\mathcal{Y}$ entries to be fetched (and therefore output entries to be allocated), would be equal to $3 + 0 + 2 + 2 = 7$. For this sub-operation, a maximum of 9 slots is needed for the accumulation probing hash table and there is no need for dynamic allocation of additional memory. 

\subsection{Summary of Performance Improvement}
 The approach described has the following time complexity, which represents an improvement over that of Sparta. The input processing stage is linear with respect to the number of nonzero entries in tensor $\mathcal{X}$ and all remaining stages have unaltered time complexity to Sparta. The reduced time complexity of the \this{} algorithm leads to a reduction in execution time while maintaining the correctness of the contraction results \footnote{An even faster time complexity could have been achieved by simply leaving the entries of tensor $\mathcal{X}$ alone during the input processing stage. This would, however, force the accumulation stage to be performed after combining intermediate entries to a global array, which would cause a significant decrease in practical performance.}. 
\begin{align}
	T_{Sparta} &=  T_{Processing} + T_{Contraction} + T_{Writeback} \\
    \text{where:}&\nonumber\\  
    \nonumber
    T_{Processing} &= \mathcal{O}({nnz}_X + nnz_Y) \\ \nonumber
    T_{Contraction} &= \mathcal{O}(nnz_X \times cmode_{Yavg}) \\ \nonumber
    T_{Writeback} &= \mathcal{O}(nnz_Z) 
\end{align} \label{eq:tc_this}

\section{Evaluation}
\subsection{Evaluation Methodology}
To validate the potential of our proposed method, \this{}, we evaluated the execution time and peak memory usage for sparse tensor contraction operations on input tensors with varying $NNZ$ and order. For fair comparison, both Sparta and \this{} were implemented in C and compiled using gcc with the optimization option -O3 and employed OpenMP for parallelism. Both programs used 8 threads maximum and were executed on a 11th Gen Intel(R) Core(TM) i7-11375H which runs @3.3GHz with 32GB of RAM. For execution time and peak memory measurements, two randomly generated tensor datasets were created with difference in mode length to observe the effect of change in sparsity. Additionally, we examined sparse tensor contraction operations between two tensors with a significant difference in $NNZ$ to assess how the algorithms handle load imbalance.

\begin{table}[hbt!]
	\caption{Here are the characteristics of tensors used in the evaluation section. Set 1 and 2 tensors contain entries with randomly generated coordinates and values to ensure no structure. Set 3 tensors originated from ITensor\cite{Fishman}}.
	\label{tlb:tensor-characteristics}
	\renewcommand{\arraystretch}{1.3}
	\centering
	\begin{tabular}{lll|lll}
		\toprule
		Name     & Shape     & NNZ  &  Name     & Shape     & NNZ  \\
		\midrule
		\multicolumn{6}{c}{Set 1}     \\   
		\cmidrule(r){1-6}
  A050 & $50 \times 50 \times 50 \times 50 \times 50$  & 50K &	B050 & $50 \times 50 \times 50 \times 50$  & 50K \\
            A100 & $50 \times 50 \times 50 \times 50 \times 50$  & 100K &	B100 & $50 \times 50 \times 50 \times 50$  & 100K \\
            A200 & $50 \times 50 \times 50 \times 50 \times 50$  & 200K &	B200 & $50 \times 50 \times 50 \times 50$  & 200K \\
            A300 & $50 \times 50 \times 50 \times 50 \times 50$  & 300K &	B300 & $50 \times 50 \times 50 \times 50$  & 300K \\
            A500 & $50 \times 50 \times 50 \times 50 \times 50$  & 500K &	B500 & $50 \times 50 \times 50 \times 50$  & 500K \\
		\midrule
          \multicolumn{6}{c}{Set 2}     \\   
            \cmidrule(r){1-6}
            A050 & $100 \times 100 \times 100 \times 100 \times 100$  & 50K &	B050 & $100 \times 100 \times 100 \times 100$  & 50K \\
		A100 & $100 \times 100 \times 100 \times 100 \times 100$  & 100K &	B100 & $100 \times 100 \times 100 \times 100$  & 100K \\
		A200 & $100 \times 100 \times 100 \times 100 \times 100$  & 200K &	B200 & $100 \times 100 \times 100 \times 100$  & 200K \\
		A300 & $100 \times 100 \times 100 \times 100 \times 100$  & 300K &	B300 & $100 \times 100 \times 100 \times 100$  & 300K \\
		A500 & $100 \times 100 \times 100 \times 100 \times 100$  & 500K &	B500 & $100 \times 100 \times 100 \times 100$  & 500K \\
		\midrule
		\multicolumn{6}{c}{Set 3}     \\               
		\cmidrule(r){1-6}
		A2190 & $4 \times 110 \times 4 \times 36 \times 486 $  & 198K &	B2190 & $36 \times 24 \times 4 \times 4$  & 81 \\
		A2178 & $4 \times 131 \times 413 \times 36 \times 4 $  & 162K &	B2178 & $36 \times 24 \times 4 \times 4$  & 81 \\
		A2177 & $4 \times 4 \times 131 \times 24 \times 413 $  & 134K &	B2177 & $24 \times 36 \times 4 \times 4$  & 95 \\
		A2164 & $131 \times 4 \times 413 \times 36 \times 4 $  & 157K &	B2164 & $36 \times 24 \times 4 \times 4$  & 81 \\
		A2163 & $4 \times 131 \times 4 \times 24 \times 413 $  & 130K &	B2163 & $24 \times 36 \times 4 \times 4$  & 95 \\
		\bottomrule
	\end{tabular}
\end{table}

\begin{table}[hbt!]
	\caption{Here are the tensor specifications for scientific applications. Tensor shapes and $NNZ$ were gathered from cited sources, tensors, besides NIPS, have randomly generated elements because specific tensor representations were not found. }
	\label{tlb:application_charaterisitics}
	\renewcommand{\arraystretch}{1.3}
	\centering
	\begin{tabular}{lllll}
		\toprule
		Name  & Order   & Shape     & NNZ  &  Density  \\
		\midrule
        \cmidrule(r){1-5}
  Enron \cite{Enron} & 3 & $186 \times 186 \times 44$  & 9838 & $6.5\times10^{-3}$ \\
  LBNL \cite{LBNL} & 3 & $65K \times 65K \times 65K$  & 27269 & $3.7\times10^{-15}$ \\
    Facebook \cite{facebook} & 3 & $63891 \times 63890 \times 1847$  & 738K & $9.8\times10^{-8}$ \\
  Hubbard-1D-P \cite{Esslinger_2010} & 5 & $4 \times 4 \times 93 \times 36 \times 432$  & 300K & $6.3\times10^{-3}$ \\
  Hubbard-1D-T \cite{Esslinger_2010}& 5 & $131 \times 4 \times 413 \times 36 \times 4$ &  400K & $5.1\times10^{-3}$ \\
  Hubbard-1D-Z \cite{Esslinger_2010}& 5 & $4 \times 129 \times 184 \times 24 \times 4$  & 100K & $5.2\times10^{-3}$ \\
  Hubbard-2D   \cite{Esslinger_2010}& 5 & $4 \times 4 \times 111 \times 24 \times 528$  & 300K & $6.6\times10^{-3}$ \\
  NIPS \cite{sparta}        & 4 & $2K \times 3K \times 14K \times 17K$  & 3M & $1.8\times10^{-6}$ \\
		\bottomrule
	\end{tabular}
\end{table}

\begin{figure}[hbt!]
	\centering
	\includegraphics[width=0.85\textwidth]{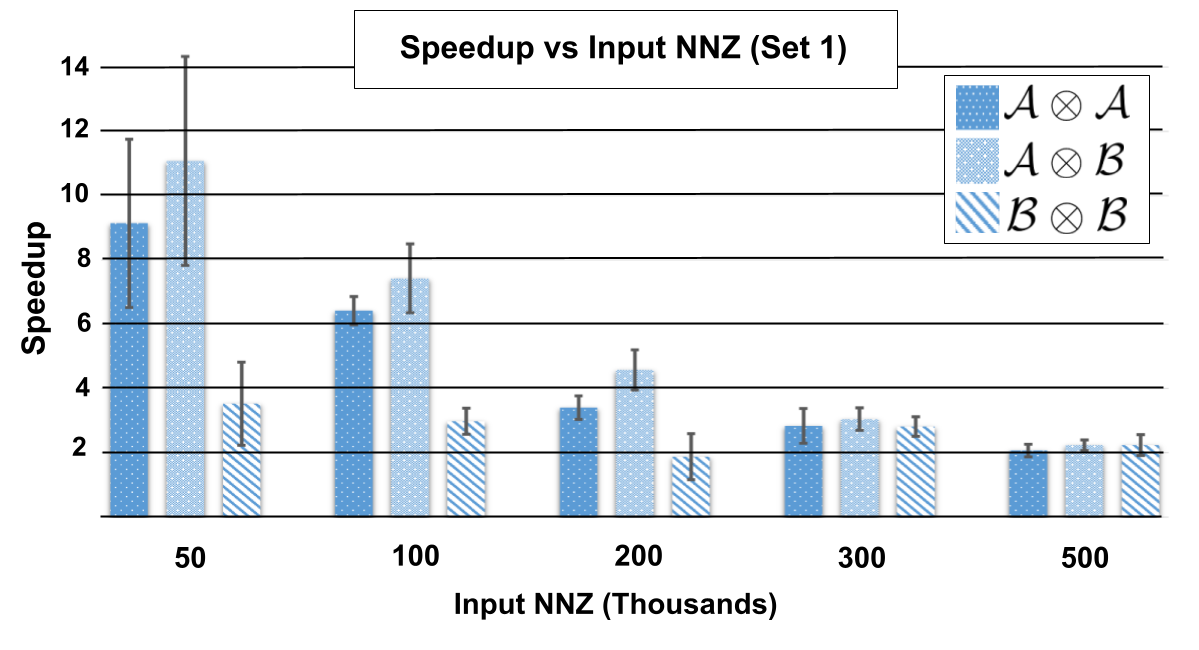}
	\caption{Speedup in contraction time versus input tensor entry counts. The specifications of the tensors used in the operations can be found in Table \ref{tlb:tensor-characteristics} Set 1. Each bar represents the average of 10 samples where each sample is an execution time of Sparta divided by an execution time of \this{}. Error bars show the standard deviation of samples. Contractions were performed along the second and third mode. }
	\label{fig:set1measurements}
\end{figure}

\begin{figure}[hbt!]
	\centering
	\includegraphics[width=0.95\textwidth]{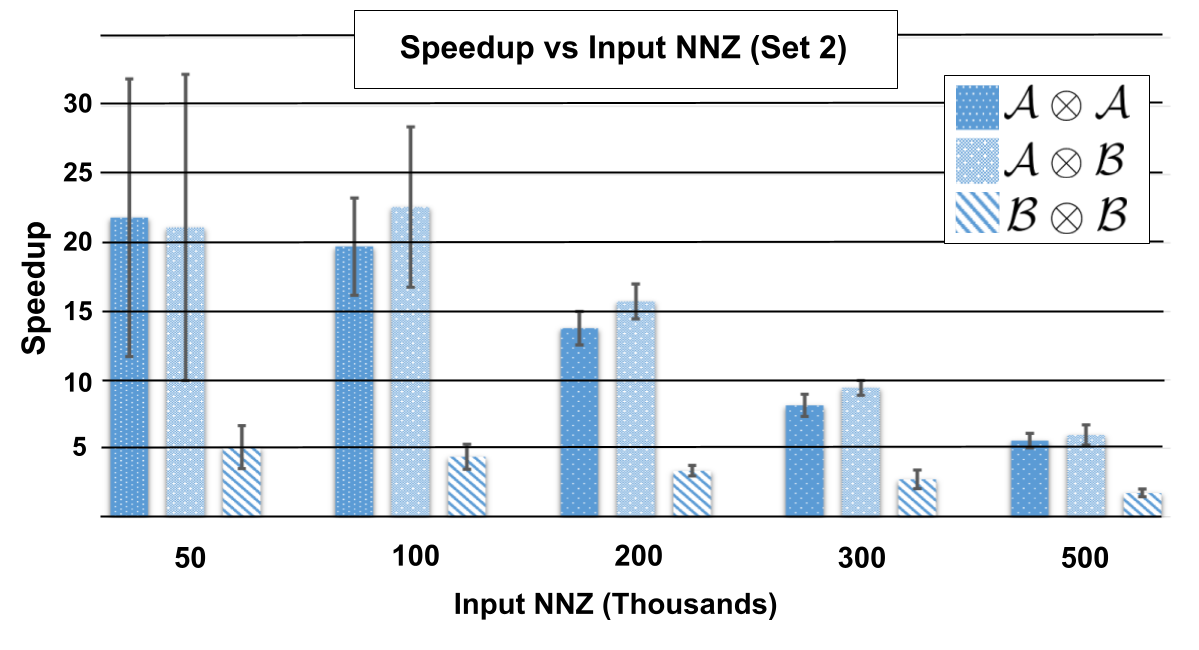}
	\caption{Speedup in contraction time versus input tensor entry counts. The specifications of the tensors used in the operations can be found in Table \ref{tlb:tensor-characteristics} Set 2. Each bar represents the average of 10 samples where each sample is an execution time of Sparta divided by an execution time of \this{}. Error bars show the standard deviation of samples. Contractions were performed along the second and third mode.}
	\label{fig:set2measurements}
\end{figure}

Figure \ref{fig:set1measurements} shows speedup in contraction time over the state-of-the-art as a function of the number of nonzero elements in the input tensors. The specifications of the tensors used in the operations can be seen in Table \ref{tlb:tensor-characteristics}, Set 1, where tensors labeled $\mathcal{A}$ are of order 5 and tensors labeled $\mathcal{B}$ are of order 4. Testing different order combinations was done to show that both designs are flexible to shape. For these measurements, both $\mathcal{A}$ \& $\mathcal{A}$ have all modes of equal length 50. Figure \ref{fig:set2measurements} similarly shows speedup as a function of $NNZ$ using the tensors described in Table \ref{tlb:tensor-characteristics}, Set 2, where modes are of equal length 100.  Tensors have randomly generated entries and values to ensure that there is no general structure that would bias the results.

\subsection{Contraction Time Versus Input Entry Counts}\label{section:speedup}

For Set 1, the minimum speedup from employing Swift sits at a solid 1.8x for large $NNZ$ with small orders to a maximum of a blistering 7.7x observed speedup for small $NNZ$ with higher orders. The speedup is much more extreme for experiments involving tensors from Set 2 which are orders of magnitude more sparse. The fastest speedup is 19.6x for small sparse tensors, and the lowest speedup is still a substantial factor of 2.3x for more dense operations.

This shows the effect sparsity can have on speedup, regardless if the $NNZ$ of the input tensors is constant. Note the formulas for time complexity of Sparta, Equation \ref{eq:tc_sparta}, and \this{}, Equation \ref{eq:tc_this}, both include the term $cmode_{Yavg}$ which represents the average number of Y entries assigned the same contracting modes. As sparsity increases, $cmode_{Yavg}$ will decrease which means a proportionately more amount of time is spent fetching entries from memory compared to processing them. Sparta and \this{} perform the same amount of time processing entries, but \this{} has much faster rate of fetching entries because of its cache-friendly storage of elements inside an array compared to a chaining hash-table. Keeping $NNZ$ and mode length constant, decreasing the order by 1 will have profound increase in density, thus self-contractions between $\mathcal{B}$ experience a lesser speedup. For reference,the sparsity for tensors from Set 1 range from 0.016\% to 8\% while tensors from Set 2 exhibit sparsity in the range of 0.0005\% to 0.5\%. 

There is a visible decrease in speedup as the number of nonzero entries increases, despite having similar time complexities. As the number of output entries increases, most of the computation time will be concentrated in the Accumulation and Writeback stages, which are comparable between the two algorithms. Consequently, less time will be spent on the Input Processing and Contraction stages, where the primary differences between the algorithms lie. There are, however, many scientific applications that utilize tensors which fall into this range of $NNZ$. In such cases, \this{} would provide significant speedup.

\begin{figure}[hbt!]
	\centering
	\includegraphics[width=0.85\textwidth]{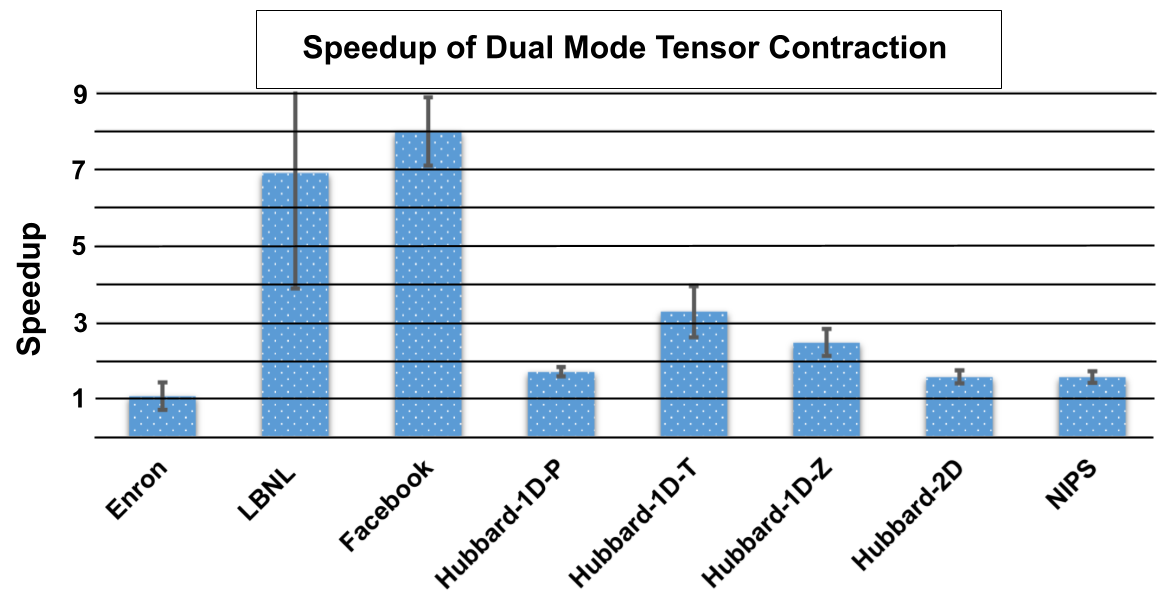}
	\caption{Execution time of SOTA divided by that of \this{} for tensors used in scientific applications.  }
	\label{fig:scientific_applications}
\end{figure}

Figure  \ref{fig:scientific_applications} shows contractions between tensors found in scientific applications. Tensor specifications for these experiments can be found in Table \ref{tlb:application_charaterisitics}. These examples further illustrate the strength of \this{} when tensors are extremely sparse. When tensors have low density, a greater proportion of execution time will be devoted towards fetching entries compared to performing the contraction operation on the entries. Tensor LBNL has a significantly lower density ($3.7\times10^{-15}$) which means modes being fetched from memory will have very few elements and thus fetching must happen more often. \this{} utilizes an array to store elements which has much faster access time compared to a chaining hash table. Tensors with such few elements, such as Enron, have little room for improvement because there are so few operations being performed and thus a reduced time complexity will have minimal impact.

\subsection{Memory Consumption Comparison}
\begin{figure}[hbt!]
	\centering
	\includegraphics[width=0.9\textwidth]{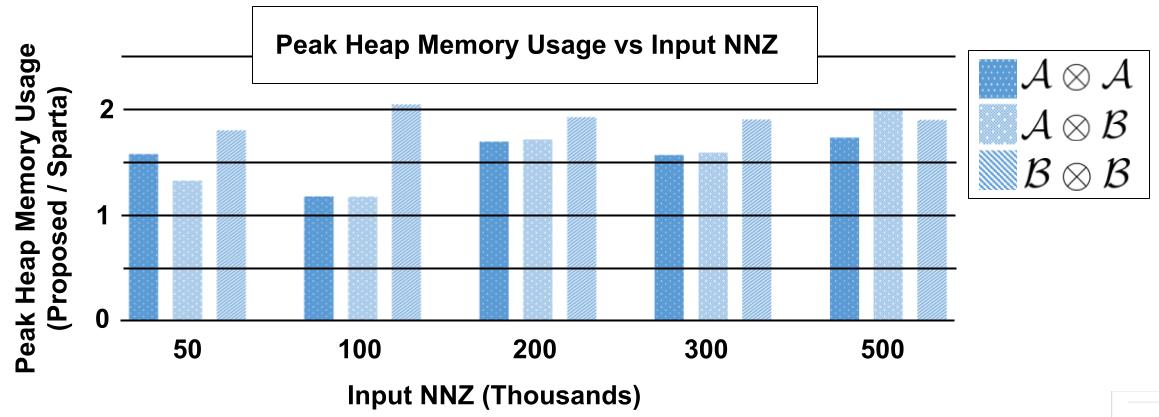}
	\caption{Ratio of heap peak memory usage for tensor contraction combinations found in Table \ref{tlb:tensor-characteristics} Set 1. Measurements show the peak memory consumption of the proposed architecture divided by that of the state-of-the-art. Contractions were performed along the second and third mode.  }
	\label{fig:memory_50}
\end{figure}
\begin{figure}[hbt!]
	\centering
	\includegraphics[width=0.9\textwidth]{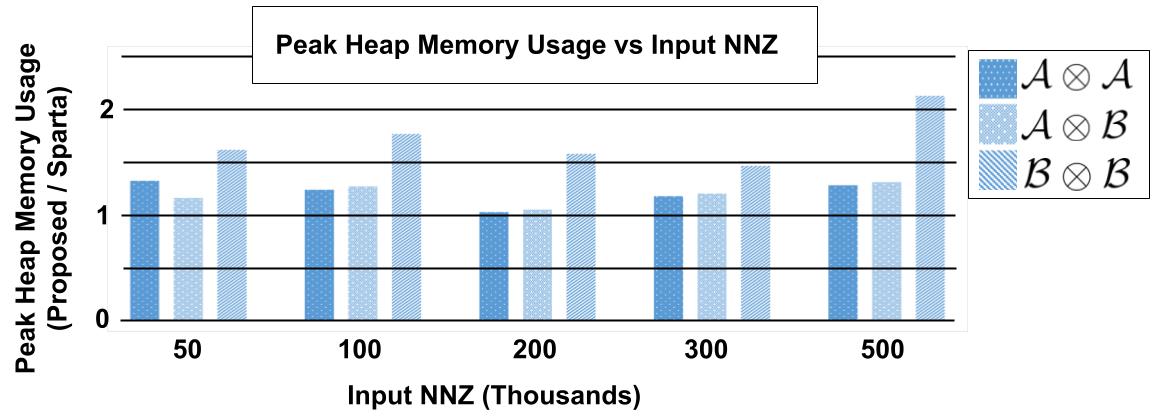}
	\caption{Ratio of heap peak memory usage for tensor contraction combinations found in Table \ref{tlb:tensor-characteristics} Set 2. Measurements show the peak memory consumption of the proposed architecture divided by that of the state-of-the-art. Contractions were performed along the second and third mode. }
	\label{fig:memory_100}
\end{figure}

For each of the experiments performed in Section \ref{section:speedup}, the peak heap memory consumption was also measured using  \texttt{memusage} on the executable files. The ratio of memory consumption from the proposed algorithm divided by that of Sparta is shown in Figures \ref{fig:memory_50} and \ref{fig:memory_100}, for tensor contraction combinations found in Table \ref{tlb:tensor-characteristics} Set 1 and 2, respectively. There is a stable increase in memory consumption as the $NNZ$ for input tensors increases, where tensors contractions between lower order tensors result in a higher ratio of memory being used. Factors which cause \this{} to require more memory are 1) the added array needed to store the counts of each contracting mode set for tensor $\mathcal{Y}$ during the contraction phase, and 2) the use of upper bounds for the accumulation hash table compared to the bare minimum being used by a chaining hash table in Sparta's accumulation stage. 
The peak memory allocations to the stack from both programs is constant with respect to the number of nonzero entries, both allocating a negligible 1.9KB peak to the stack. 


\subsection{Imbalanced Contractions}
\begin{figure}[hbt!]
	\centering
	\includegraphics[width=1\textwidth]{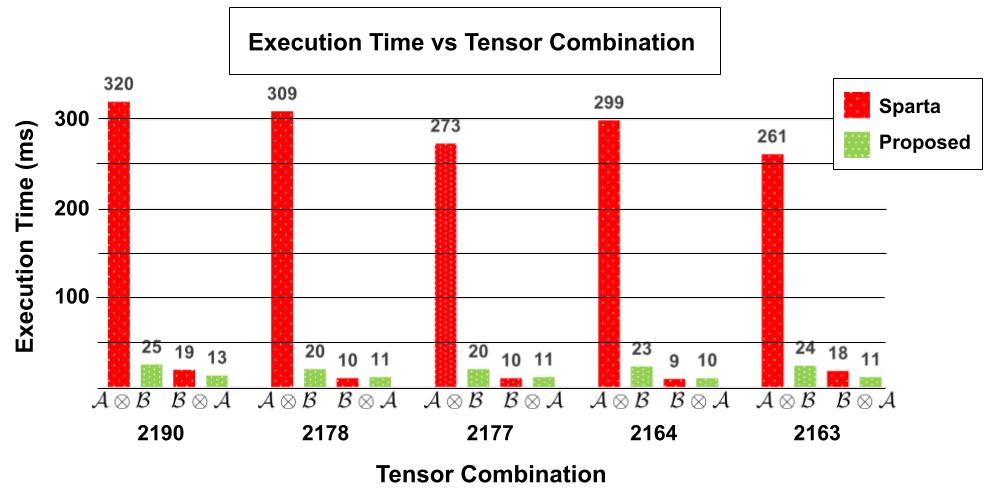}
	\caption{Tensor contraction time in milliseconds for various different combinations of tensors. See Table \ref{tlb:tensor-characteristics} Set 3 for specifications of tensors used in the contraction operations. }
	\label{fig:imbalancedmeasurements}
\end{figure}

The third set of tensor contractions in Table \ref{tlb:tensor-characteristics} contains imbalanced operations in which the left tensor has vastly more non-zero elements compared to the right. Although both operations result in the same output tensor, this section compares the difference in contraction time when the order of the tensors in the operation are swapped. This can be important in the case of unknown tensor sizes such as the instance these algorithms are used in a chain of contractions where the output tensor of the previous layer has an unknown number of non-zero elements. The measurements are displayed in Figure \ref{fig:imbalancedmeasurements}. As can be seen, the operation $\mathcal{B} \otimes \mathcal{A}$ is similar for both algorithms, but the operation $\mathcal{A} \otimes \mathcal{B}$ is very imbalanced and Sparta has much worse performance. This imbalance should be due to the imbalanced time complexity and the difference in tensor representations based on the order of contraction for Sparta, versus \this{} which represents both tensors in the same fashion and has a symmetric time complexity. For these extremely imbalanced cases, the Sparta execution time increases by a factor of 26x on average when the order of the operations is swapped, and \this{} execution time increases by a factor of 0.96x when the order of operations is swapped. 

\section{Conclusion}
Sparse Tensor Contraction is important for a large range of applications that work with higher dimensional data. There are numerous challenges, the most prominent of which are irregular memory access patterns, unknown output tensor size, and high amounts of intermediate data. The state-of-the-art handles these problems through the use of hash table based representation of tensor $\mathcal{Y}$, sorting tensor $\mathcal{X}$, and using chaining hash tables to place intermediate entries into a dynamic array. This work presents alterations to key stages that are shown to improve both the performance of the operation and time complexity compared to the state-of-the-art sparse tensor contraction algorithm. The most important changes are the grouping of tensor $\mathcal{X}$ to improve performance and time complexity, representing tensor $\mathcal{Y}$ as an array with pointers to divisions, and using a probing hash table at the output of accumulation for better memory access time. These improvements are shown to have up to 20$\times$ speedup over diverse test cases with a trade-off of slightly increased memory allocation.

\section{Acknowledgment}
This research was supported, in part, by the National Science Foundation grants 2217028 and 2316203.
\input{sn-article.bbl}

\end{document}